# Haptic User Interfaces and Practice-based Learning for Minimally Invasive Surgical Training

**Felix G. Hamza-Lup[1,2], Adrian Seitan[2], Costin Petre[2], Mihai Polceanu[2], Crenguta M. Bogdan[2], Dorin M. Popovici[2]**

Computer Science, Armstrong Atlantic State University,
11935 Abercorn St., Savannah, GA, USA
Mathematics and Informatics, Ovidius University
124 Mamaia Blvd., Constanta, ROMANIA
E-mail: felix.hamza-lup@armstrong.edu

Abstract

*Recent advances in haptic hardware and software technology have generated interest in novel, multimodal interfaces based on the sense of touch. Such interfaces have the potential to revolutionize the way we think about human-computer interaction and open new possibilities for simulation and training in a variety of fields.*

*In this paper we review several frameworks, APIs and toolkits for haptic user interface development. We explore these software components focusing on minimally invasive surgical simulation systems.*

*In the area of medical diagnosis, there is a strong need to determine mechanical properties of biological tissue for both histological and pathological considerations. Therefore we focus on the development of affordable visuo-haptic simulators to improve practice-based education in this area. We envision such systems, designed for the next generations of learners that enhance their knowledge in connection with real-life situations while they train in mandatory safety conditions.*

## 1 Introduction

Computer based simulation and training environments are primarily relying on the visual and auditory senses for human-computer interaction. The past decade has seen an accelerated evolution of both hardware and software, specifically hardware capable of simulating touch (i.e. haptics).

Haptics is the science of merging tactile sensation with computer applications, thereby enabling users to receive feedback they can feel in addition to auditory and visual cues. Multimodal environments where visual, auditory and haptic stimuli are present convey information more efficiently since the user manipulates and experiences the environment through multiple sensory channels (Hamza-Lup, 2009). Such environments were proposed, and analysed in the past decade in conjunction with medical training, specifically minimally invasive surgery. In the area of medical diagnosis, there is a strong need to determine mechanical properties of biological tissue for both histological and pathological considerations. One of the established diagnosis procedures is the palpation of body organs and tissue. Therefore we focus on the development of affordable visuo-haptic simulators to improve practice-based education in this area by proposing a simulation system for liver palpation.

The paper is structured as follows. Section 2 presents a brief review of the practice-based learning paradigm followed by the visuo-haptic user interface architecture in Section 3. In this section we also present a brief description of the main frameworks and APIs for Haptic User Interface (HUI) development. Section 4 illustrates the use of two APIs H3D and CHAI3D in the development of a HUI for a haptic liver palpation system. We conclude in Section 5 with remarks useful in the development of HUI for laparoscopic surgical simulators.

## 2   Practice-based Learning

Practice-based learning is best understood in contrast to "theory-based" learning. It is related to terms such as "work-based" or '"work-centred" learning. Practice-based learning main focus is the formation of effective and self-renewing practices, often backed by a solid theoretical foundation. Practice-based approaches potentiate understanding and knowledge retention in learning and training settings.

Recent trends related both to higher education and healthcare delivery systems have created an environment of change for medical education. Educators, physicians and administrators face a number of challenges, such as the introduction of more complicated technology with which to deliver health care, limitations on orientation times, a need to address concerns related to patient safety and quality of care - all of which require new and creative solutions.

One solution to the growing concerns linked to these trends is the advent of health care simulation and training in medical education programs and in hospitals. These simulations, which serve as adjuncts to didactic learning, represent the closest possible technology to real patients and allow for a repetitive "hands-on" learning in a safe environment where mistakes can be safely made. The students who participate in simulations gain experience and confidence on their ability to make critical clinical decisions in acute care situations, where time and skill have critical consequences.

The practice of medicine has always relied on visualizations. These visualizations have either been direct or have required extensive mental reconstruction. The revolutionary capabilities of new three-dimensional (3D) and four-dimensional (4D) imagining modalities underscore the vital importance of spatial visualization to this science (Knottnerus et al., 2008).

The introduction of the sense of touch through haptic technology in simulations opens new realms/perspectives in the endeavour of teaching and practicing medicine, specifically in surgical simulation. Depending on the specific task, force feedback can provide benefits (both passive and informational) at little cost to cognitive demand because of its intuitive nature (Wagner et al 2005).

## 3   Haptic User Interface Architecture

HUIs rely on two components, the visual component (i.e. 3D objects and their visual behaviour) and the haptic behaviour associated with the elements of the scene. The main software components of a HUI for a virtual environment simulating surgical procedures are illustrated in Figure 1:

- Multimodal input (mouse, keyboard, haptic device)
- Multimodal output (3D visualization system, haptic device)
- Simulation engine (rendering cycle per modality + synchronization)
- Environment persistency (database).

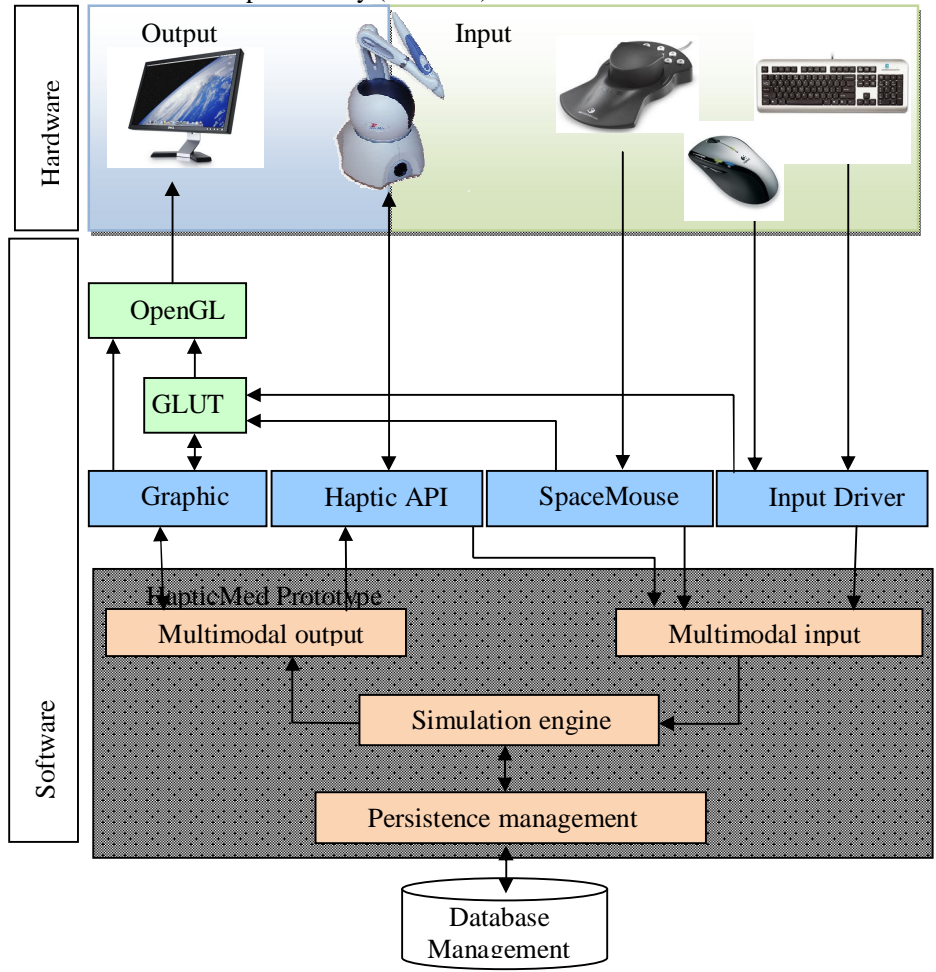

**Figure 1. HUI the interconnections among the software components and the hardware components involved**

The multimodal input component controls the user interaction devices: the keyboard, mouse, 3D navigation (SpaceMouse) and haptic device (Phantom Omni). It allows the user to express his/her actions in the virtual environemnt. The multimodal output component conveys information to the user through multiple sense: visual – the 3D visualization system, tactile – force feedback felt through the Phantom Omni device and possibly audio using external speakers.

The simulation engine consists of several cycles that interleave to generate the multimodal feedback: a graphic rendering cycle and a haptic rendering cycle (and possibly an audio rendering cycle).

The simulation engine receives the user input data from the devices through the software components associated with each device. The simulation engine will compute the behaviour and appearance of each component in the virtual environment based on specific laws of physics, collisions, providing realistic behaviour as an effect to user's actions. Specifically for surgical simulations, the tissue deformation dynamics must be accurately simulated both visually and tactile. The graphic computational cycle varies between 20 to 60 frames per second while the haptic cycle must be kept at 1000 frames per second (1KHz). The cycles must be synchronized to assure correct visual and haptic rendering. The simulation complexity as well as the surgical task accuracy requires a careful balance and tradeoff among geometric complexity, deformation accuracy and execution speed.

The simulation engine uses the Database Management System (DBMS) to store/retrieve specific elements used in the description of the virtual world. The DBMS enables actions like object geometry and object properties memorization.

### 3.1 The Visual Component - the Scene Graph

One of the best known libraries for computer graphics development is undoubtedly OpenGL (OpenGL, 2011). OpenGL is a low level graphics library that allows development of 2D and 3D graphics using low level primitive like vertices, lines and polygons. A competitor for OpenGL is DirectX. Microsoft DirectX is a collection of application programming interfaces (APIs) for handling tasks related to multimedia, especially game programming and video, on Microsoft platforms.

While both standards have been used in computer graphics project development, new higher level standards and API have emerged in the past decade. One of the most proeminent example is the successor of VRML (Virtual Reality Modeling Language), X3D (Extended 3D Graphics) (X3D, 2011). X3D is a royalty-free open standards file format and run-time architecture to represent and communicate 3D scenes and objects using XML.

In virtual environments the visual component is usually quite complex. In order to cope with such complexity a specific data structure is necessary. A scene graph is a general data structure commonly used by vector-based graphics. The scene graph is a structure that arranges the logical and often spatial representation of a graphical scene.

A scene graph is a collection of nodes in a graph or tree structure. A node may have many children but often only a single parent, with the effect of a parent applied to all its child nodes; an operation performed on a group automatically propagates its effect to all of its members. A common functionality is the ability to group related shapes/objects into a compound object that can then be moved, transformed, selected, etc. as easily as a single object.

This scene graph is augmented with tactile attributes corresponding to virtual objects that the user manipulates inside the virtual environment. To employ the haptic paradigm in the simulation one needs to use a haptic-oriented framework.

### 3.2 The Haptic Component – Toolkits, APIs and Frameworks

In recent years as the haptic (force feedback) paradigm became more popular, several APIs and later frameworks have surfaced. We explored the most actual haptic frameworks and APIs and compared their features and capabilities. In the following paragraphs we provide a brief description of each framework and API investigated along with their main architecture.

#### 3.2.1 OpenHaptics Toolkit

The OpenHaptics toolkit (OpenHaptics, 2011) developed by SenseAble Technologies includes the QuickHaptics interface, the haptic device interface (HDAPI), the haptic library interface (HLAPI) as well as utilities and drivers for the Phantom haptic device family. The toolkit is backed by a solid documentation and a programmer's guide. The toolkit architecture is illustrated in Figure 2.

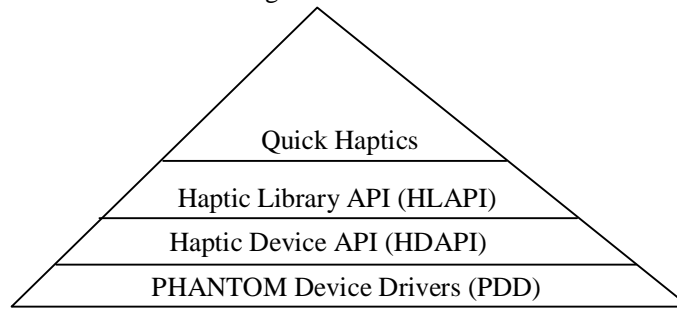

**Figure 2. OpenHaptics toolkit - architecture**

The HDAPI provides low level access to the haptic devices. The programmer can replay forces on the device and has access to the device driver configuration settings and debugging support. The HLAPI covers haptic feedback at a higher level and requires OpenGL development knowledge. QuickHaptics allows haptic application development or extensions for existing applications.

#### 3.2.2 ReachIn API

The ReachIn API (Reachin, 2009) is a modern development platform that enables the development of sophisticated haptic 3D applications in the user's programming language of choice, such as C++, Python, or VRML. The API provides a base of pre-written code that allows easy and rapid development of applications that target the specific user's needs. UK Haptics (UK Haptics, 2011), a recently established medical software development company, used ReachIn API as the core haptic technology platform for their Virtual Veins project, a medical simulation package for training medical staff in catheter insertion.

The ReachIn API (illustrated in Figure 3) is organized as follows:
– External device interfaces: haptic, graphic, audio and other non-haptic devices.
– The rendering engine synchronizes three rendering cycles: haptic, graphic and audio or a cycle without haptic devices.

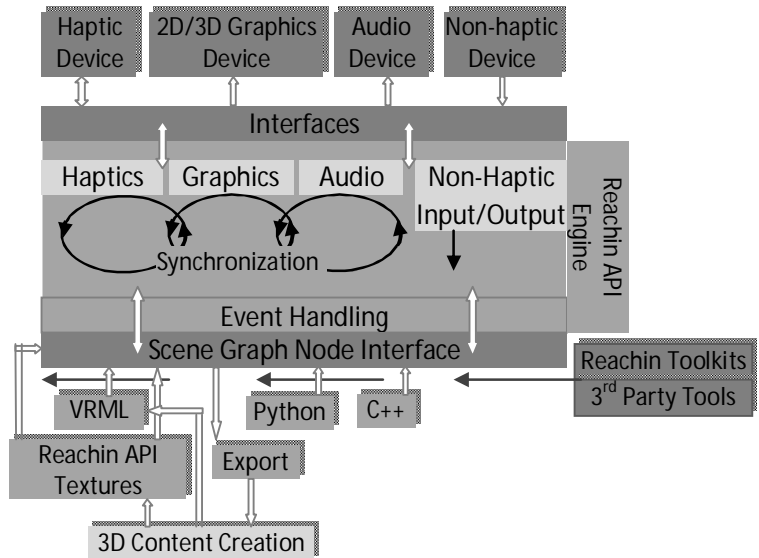
**Figure 3. ReachIn API – architecture**

The simulation engine follows the scene graph to find and execute the graphic and haptic routines based on the scene structure. The node is the structural unit and stores the visual and haptic properties for each object in the scene.

### *3.2.3 Haptics 3D API*
Haptics3D (H3D, 2011) is one of the best-known open source APIs that bridges OpenGL and X3D with haptic components. Figure 4 illustrates the architecture of H3D.

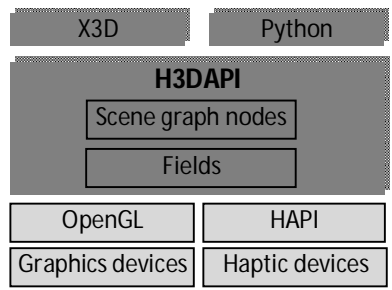
**Figure 4. H3D API - Architecture**

This API is designed mainly for users who want to develop haptic-based applications from scratch, rather than for those who want to add haptics to existing applications. The main advantages of H3D are the rapid prototyping capability and the compatibility with X3D, making it easy for the developer to manage both the 3D graphics and the haptic rendering. H3D API uses the X3D and OpenGL standards and builds on haptic technology from SensAble's OpenHaptics™ toolkit (OpenHaptics, 2011). It allows users to focus their work on the behaviour of the application, and ignore the issues related to haptics geometry rendering. The API is also extended with scripting capabilities, allowing the user to perform rapid prototyping using the Python scripting language.

### *3.2.4  Computer Haptics and Active Interfaces Framework*

Developed with medical applications in mind, the Computer Haptics & Active Interfaces (CHAI 3D) (CHAI3D, 2011) is an open source set of C++ libraries supporting haptic-based systems, visualization, and interactive real-time simulation. The API facilitates the integration of 3D modeling with haptic rendering. Moreover, the applications are portable and can be executed on different platforms.  This quality attribute is obtained by saving object characteristics in XML files. The applications can be tested using a real haptic device (e.g., PHANToM Omni), or a virtual representation using the mouse as a substitute for the haptic device. The API was recently extended with a simulation engine for rigid/deformable objects.

CHAI3D supports multiple commercial interfaces (e.g. IEEE 1394, Servo2Go, Sensoray626) facilitating haptic device connection with the application. The framework can be extended using ODE modules.  (ODE, 2001) is an open source, high performance library for simulating rigid body dynamics. It is fully featured, stable, mature and platform independent with an easy to use C/C++ API. It has advanced joint types and integrated collision detection with friction.

CHAI3D architecture is organized as a set of packages (as illustrated in Figure 5) that contain classes and interfaces for 3D scene development and for the application of material properties on the objects in the scene. For example, the InPort package contains classes that support the I/O device communication e.g. mouse, joystick, glove and the Phantom Omni haptic device.

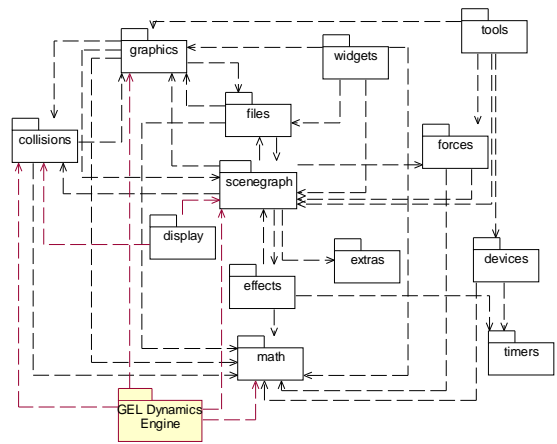

**Figure 5. CHAI3D framework – architecture**

### *3.2.5  Simulation Open Framework Architecture*

The need for standardization and inter-project cooperation gave rise to the Simulation Open Framework Architecture (SOFA, 2011). SOFA is targeted at real-time simulation, with an emphasis on medical simulation. It allows the development of multiple geometrical models and the simulation of the dynamics of interacting objects using abstract equation solvers. An additional advantage of this framework is the use of the XML standard to streamline the parameters of the simulation like deformable behaviour, collision algorithms, and surface constraints.

Objects representation is represented through several models: a dynamic model, where the object mass and composition are specified, a collision model, used to compute the object collision behaviour, and a visual model, containing the object's geometry (Figure 6).

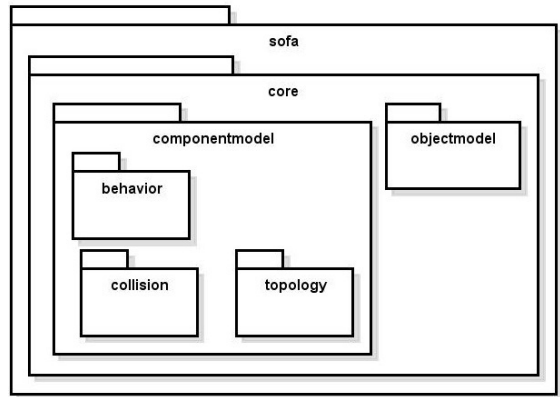

**Figure 6. SOFA framework – architecture**

### 3.2.6 General Physical Simulation Interface

Another effort targeted at applications of haptics in surgical simulators is the General Physical Simulation Interface (GiPSi, 2011). It is a general open source/open architecture framework for developing organ level surgical simulations. The framework provides an API for interfacing dynamic models defined over spatial domains. It is specifically designed to be independent of the specifics of the modeling methods used and therefore facilitates seamless integration of heterogeneous models and processes. The framework contains I/O interfaces for visualization and haptics integration in applications, a set of modeling and computational tools and a simulation kernel as illustrated in Figure 7.

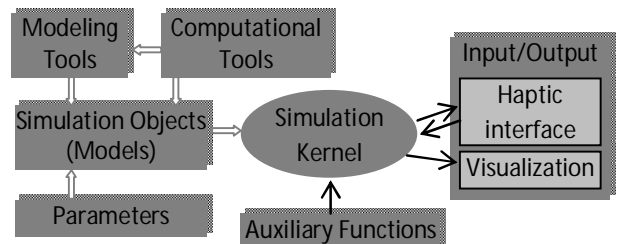

**Figure 7. GiPSi framework - architecture**

GiPSi is known for the simulation of internal organs specifically the heart (Cavusoglu, 2004).

In the context of the HapticMed project (HapticMed, 2011) we are focusing on the development of a set of simulation tools for minimally invasive surgical training. Laparoscopic procedures are frequently used in surgical treatment of liver diseases.

## 4 Haptics@work – Liver Diagnosis through Palpation

The liver is the largest organ in the human body. During development, liver size increases with increasing age, averaging five centimetres span at five years and attaining adult size by age fifteen. The size depends on several factors: age, sex, body size and shape, as well as the particular examination technique utilized (e.g., palpation versus percussion versus radiographic). By percussion, the mean liver size is seven centimetres for women and about ten centimetres for men. A liver span two to three centimetres larger or smaller than these values is considered abnormal. The liver weighs 1200 to 1400 g in the adult woman and 1400 to 1500 g in the adult man (Gilbert, 1994).

The normal liver is smooth, with no irregularities. When the liver can be felt, it is usually due to:
- increased diaphragmatic descent;
- presence of a palpable caudate or Riedel's lobe;
- presence of emphysema with an associated depressed diaphragm;
- thin body habitus with narrow thoracic cage;
- fatty infiltration (enlarged with rounded edge);
- active hepatitis (enlarged and tender);
- cirrhosis (enlarged with nodular irregularity);
- hepatic neoplasm (enlarged with rock-hard or nodular consistency).

In laparoscopic surgery internal tissue palpation is an important pre-operatory activity (Kim et al, 2004; Khaled, 2004). The normal liver may be slightly tender upon palpation, but the inflamed liver (hepatitis) is often exquisitely tender. The nodularity, irregularity, firmness, and hardness of the liver can be characterized.

In the framework of the HapticMed project we have designed a visuo-haptic prototype for liver palpation. We have implemented a 3D deformable model of the liver. The simulator replicates several disease conditions that must be recognized by the user. The final goal is to provide an advanced testing mechanism to assess medical student knowledge related to liver pathology. The simulator was implemented using H3D API as well as the CHAI3D framework to provide us with additional information regarding ease of development using different software systems.

### 4.1 Prototype Implementation with the H3D API

The H3D implementation relies on a deformable mesh model. The liver 3D model uses the X3D (X3D, 2011) triangle set. We started with a 40.000 polygonal model and we reduced it to 3.200 in order to obtain real-time deformation behaviour. The deformation algorithm is a classic spring-damper, meaning that the haptic device will render a force directly proportional with the penetration distance into the deformable object (illustrated in Figure 8).

Two important parameters when implementing a deformable 3D object is the object's surface type and its stiffness. The first one defines the visual and haptic properties at touch while the second one is part of the deformation algorithm that provides visual as well as force feedback during object palpation and/or penetration.

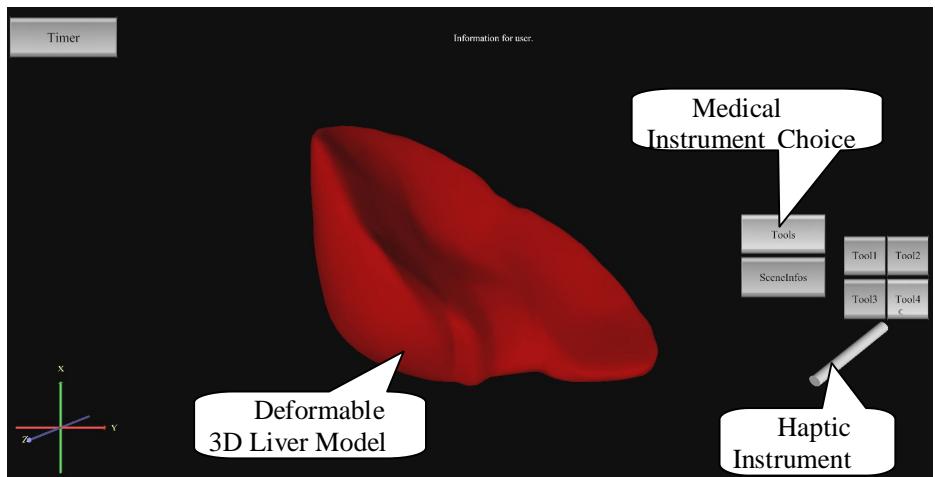
**Figure 8. HUI live palpation simulator with H3D**

The H3D API provides an easy solution for visuo-haptic deformation through the use of the *DeformableShape* node that uses two kinds of geometries: a haptic geometry and a visual geometry.

### 4.2 Prototype Implementation with the CHAI3D Framework

CHAI3D haptic deformation we relies on the GEL framework based on a set of nodes (spheres) attached with springs as illustrated in Figure 9.

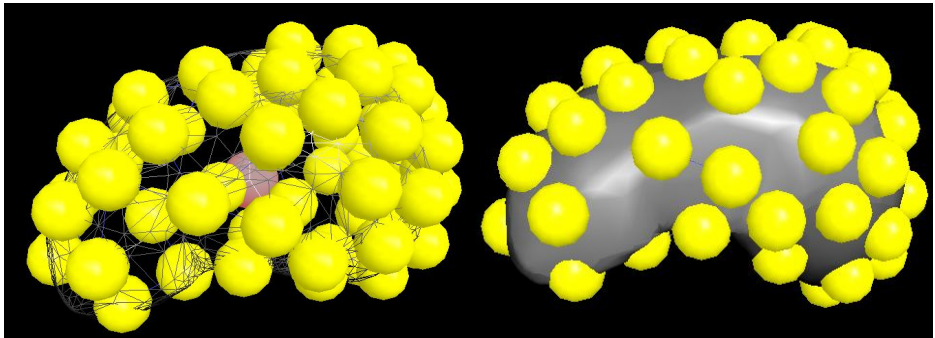
**Figure 9. HUI live palpation simulator with CHAI3D (wireframe-left)**

In this framework the real-time deformation of a complex structure is possible through spherical skeleton models. Collision detection between different objects can be effectively estimated by computing the collision between their skeletons.

For the liver dynamics simulation we employ three spring connectors for each node. The nodes are spread on the surface of the liver taking in consideration the distance between two nodes, such that we obtain a uniform distribution as illustrated in Figure 9. We are in the process of adjusting the spring model damping factor to obtain a realistic behaviour for the liver deformation and for force feedback computation.

## 5  Conclusion

The potential of haptic interfaces in support of practice based learning is proved more often in medical training. Even if the haptic hardware is more affordable, the development of haptic based simulations is hampered by the lack of existing software adaptability and extensibility.

We are in the process of developing and assessing a set of simulation tools for minimally invasive surgical training which are both useful and affordable for large scale deployment in hospitals.

### 5.1  Acknowledgments

This study was supported under the ANCS Grant "HapticMed – Using haptic interfaces in medical applications", no. 128/02.06.2010, ID/SMIS 567/12271.